\documentclass[twocolumn,prd,floatfix,preprintnumbers,a4paper,nofootinbib,superscriptaddress,groupedaddress]{revtex4-1}
\usepackage[utf8]{inputenc}
\usepackage{amsmath,amssymb}
\usepackage{amsfonts}
\usepackage{bm}
\usepackage{graphics}
\usepackage{float}
\usepackage{amsmath}
\usepackage{amssymb}
\usepackage[normalem]{ulem} 
\usepackage[mathscr]{euscript}
\usepackage{acronym}
\usepackage{float}
\usepackage[caption=false]{subfig}
\usepackage{lipsum}
\usepackage{mathtools}
\usepackage{multirow}
\usepackage{graphicx}
\usepackage{mathtools}
\usepackage{multirow}
\usepackage{graphicx}
\usepackage[usenames,dvipsnames,table,xcdraw]{xcolor}
\usepackage[colorlinks, pdfborder={0 0 0}]{hyperref} 
\usepackage[nameinlink,capitalise]{cleveref}
\usepackage{soul} 

\usepackage{enumerate}

\begin{document}


\title{Gravitational-wave detectors as particle-physics laboratories:\\Constraining scalar interactions with a coherent inspiral model of boson-star binaries}

\author{Costantino Pacilio}
\email{costantino.pacilio@uniroma1.it}
\affiliation{%
Dipartimento di Fisica, “Sapienza” Università di Roma \& Sezione INFN Roma1, Roma 00185, Italy
}%

\author{Massimo Vaglio}
\email{massimo.vaglio@uniroma1.it}
\affiliation{%
Dipartimento di Fisica, “Sapienza” Università di Roma \& Sezione INFN Roma1, Roma 00185, Italy
}%

\author{Andrea Maselli}
\email{andrea.maselli@roma1.infn.it}
\affiliation{%
Dipartimento di Fisica, “Sapienza” Università di Roma \& Sezione INFN Roma1, Roma 00185, Italy
}%

\author{Paolo Pani}
\email{paolo.pani@uniroma1.it}
\affiliation{%
Dipartimento di Fisica, “Sapienza” Università di Roma \& Sezione INFN Roma1, Roma 00185, Italy
}%

\begin{abstract}
Gravitational-wave~(GW) detections of binary neutron star coalescences play a crucial role to constrain the microscopic interaction of matter at ultrahigh density. Similarly, if boson stars exist in the universe their coalescence can be used to constrain the fundamental coupling constants of a scalar field theory. We develop the first coherent waveform model for the inspiral of boson stars with quartic interactions. The waveform includes coherently spin-induced quadrupolar and tidal-deformability contributions in terms of the masses and spins of the binary and of a single coupling constant of the theory. We show that future instruments such as the Einstein Telescope and the Laser Interferometer Space Antenna can provide strong complementary bounds on bosonic self-interactions, while the constraining power of current detectors is marginal.
\end{abstract}

\maketitle

\section{Introduction}
Gravitational-wave~(GW) measurements of the tidal deformability 
of neutron stars~(NSs) have opened a new window to study 
the properties of matter beyond the nuclear saturation point 
within stellar cores~\cite{Hinderer:2018mrj,Chatziioannou:2020pqz}. 
Equations of state with different stiffness provide tidal 
deformabilities which may vary up to an order of magnitude. This 
effect magnifies the details of the underlying microscopic model, 
allowing to probe how fundamental interactions behave in extreme 
regimes \cite{Lattimer:2006xb,Maselli:2018fay}.

In this paper, we argue that the very same situation can occur if boson stars~(BSs)~\cite{kaup1968,ruffini1969} exist in the Universe and form coalescing binaries within the horizon of current and future detectors. BSs are self-gravitating condensates of a bosonic field (see Refs.~\cite{liebling2017review,jetzer1992review} for some reviews). In their original and simplest proposal, they are solutions to Einstein gravity minimally coupled to a classical field theory for a complex scalar $\phi$:
\begin{equation}
    {\cal L}_{\rm scalar}=\frac{1}{2}\partial_\mu\phi^\star\partial^\mu\phi+V\left(\left|\phi\right|\right)\,, \label{Lagrangian}
\end{equation}
where a star denotes complex conjugation and $V$ is the scalar self potential. The latter plays the same role as the equation of state for NSs: different microscopic interactions give rise to macroscopically different properties of the boson stars. 

Depending on the mass of the boson field and on the self-interaction terms, BSs can exist in any mass range and can have a compactness comparable to or larger than that of a NS. It is intriguing that current GW measurements cannot exclude the existence of exotic compact objects other than black holes~(BHs) and NSs, especially for GW events in the low-mass~\cite{Abbott:2020khf} and high-mass gap, where neither BHs nor NSs are predicted in the standard scenario.

As a case study, in this paper we consider a simple class of quartic interactions [see Eq.~\eqref{potential:2}] and quantify the accuracy within which a GW detection of a BS coalescence can constrain the fundamental parameters (boson mass and coupling constants) of a given scalar field theory.

We focus on the inspiral phase, which can be accurately modeled with post-Newtonian~(PN) theory~\cite{blanchet2014review,PoissonWill}. Up to $1.5$ PN order (see below), the GW signal depends only on the masses and spins of the binary components and is, therefore, oblivious to the nature of the latter. However, the details of the coalescing bodies appear at higher PN order, notably through the effects of the spin-induced quadrupole moment (if the binary is spinning)~\cite{poisson1998quadrupole,PoissonWill}, a small tidal-heating term (if at least one of the binary components is a BH or can efficiently absorb radiation)~\cite{hartle1973,hughes2001,Maselli:2017cmm}, and most importantly through the tidal deformability contribution (the so-called tidal Love numbers~\cite{flanagan2008,Hinderer:2007mb,PoissonWill}) that becomes increasingly more relevant during the late stages of the inspiral and merger, as in the case of a binary NS coalescence (see, e.g., Refs.~\cite{Harry:2018hke,Chatziioannou:2020pqz} for some recent reviews).

Previous work considered the aforementioned effects independently and included in the waveform only a single effect at the time, focusing on the detectability of the tidal Love number~\cite{cardoso2017,sennett2017}, or of the spin-induced quadrupole moment~\cite{Kastha:2018bcr,krishnendu2017,krishnendu2019,krishnendu2019_2}, or of the tidal heating alone~\cite{Maselli:2017cmm,Datta:2019epe,Datta:2019euh,datta2020}. However, this approach neglects a crucial ingredient: for a given scalar field theory (i.e., fixing the potential in the Lagrangian~\eqref{Lagrangian}), both the tidal Love numbers and the spin quadrupole moment depend only on the masses and spins of the binary. Therefore, their concurrent inclusion does not increase the number of waveform parameters and alleviates their degeneracy. 

\begin{figure}[t]
    \includegraphics[width=0.48\textwidth]{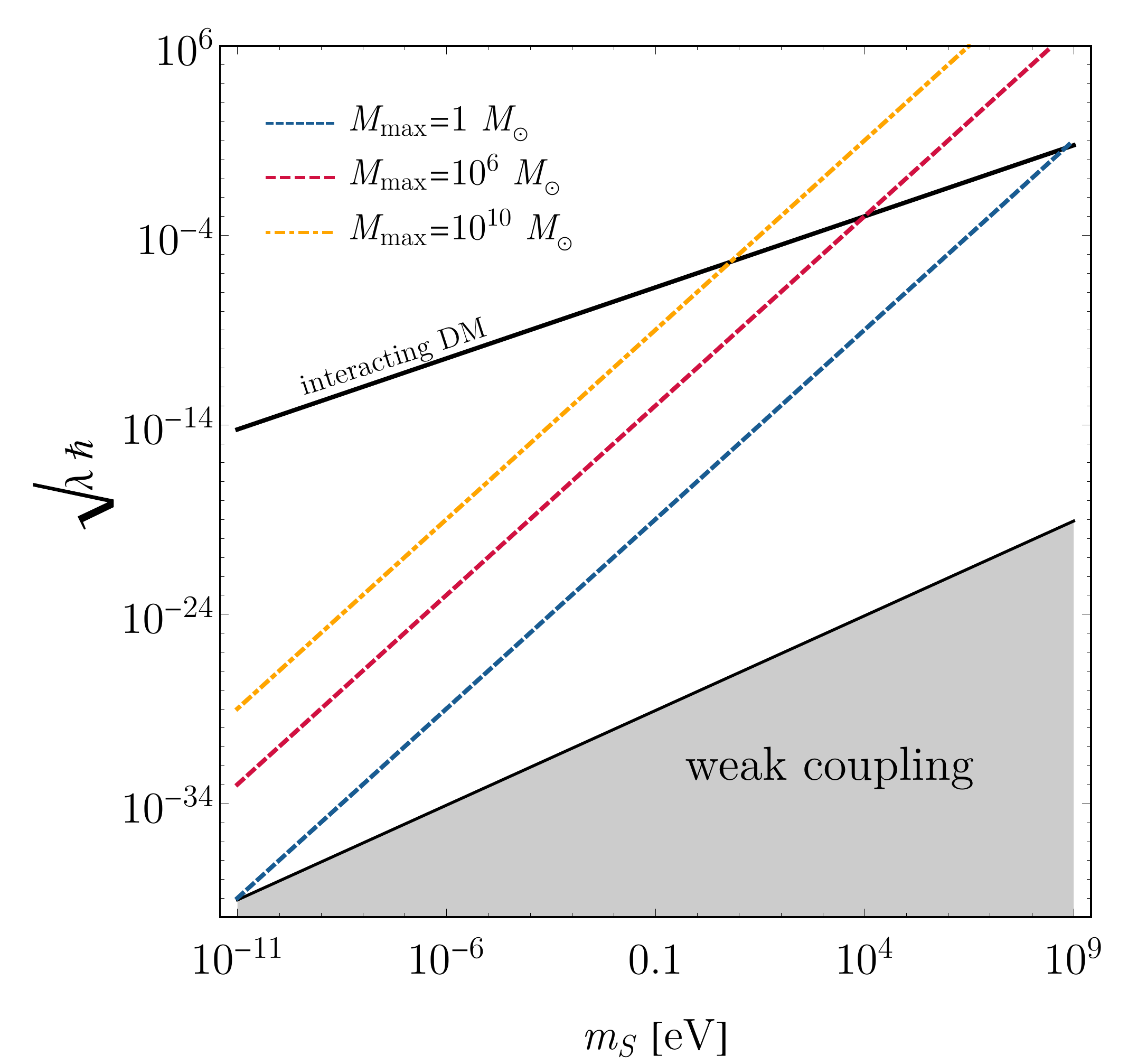}
    \caption{Parameter space of a scalar field theory with a quartic potential, Eq.~\eqref{potential:2}. The shaded area denotes the weak-coupling region, $\lambda<\mu^2$, whereas the three parallel inclined lines correspond to Eq.~\eqref{mmax:1} for different values of $M_{\rm max}$ and give an estimate of the region that can be probed by astrophysical BSs. The continuous black line corresponds to a model of self-interacting dark matter~\cite{giudice2017}, see Eq.~\eqref{selfDM}.}
    \label{fig:paramspace}
\end{figure}

As we shall show, the sensitivity of current detectors such as the Laser Interferometer Gravitational Wave Observatory (LIGO) and Virgo is not sufficient to place stringent constraints on the coupling constants of the theory. However, future facilities will provide much more stringent measurements. In particular, we consider the future Laser Interferometer Space Antenna~(LISA)~\cite{Audley:2017drz}, which can potentially detect supermassive binary BSs, and the Einstein Telescope~(ET)~\cite{Maggiore:2019uih}, a proposed third-generation~\cite{Hild:2010id} ground-based GW detector. Putative detections of binary BSs in different mass ranges can provide complementary constraints on the fundamental parameters of interacting scalar-field theories (see Fig.~\ref{fig:paramspace}), thus turning GW detectors into particle-physics laboratories~\cite{barack2019review,giudice2017}.

The plan of the work is as follows. Section~\ref{sec:boson stars} reviews the main properties of BSs in our model, in particular their mass, tidal deformability, and spin-induced quadrupole moment. Section~\ref{sec:waveform} reviews how those properties enter in the PN expansion of the waveform. Section~\ref{sec:fisher} reviews the Fisher matrix formalism to estimate statistical errors on the model parameters and discusses previous work on the topic. Section~\ref{sec:results} presents our results on the projected measurements of the fundamental coupling constant of the theory using GW detections of binary BSs. Finally, Appendix~\ref{appendix:1} summarizes our fits and useful relations between various BS parameters.

We use $G=c=1$ units, whereas we keep Planck's constant $\hbar$ explicit, defining the Planck mass as $M_P=\sqrt{\hbar}$.

\section{BS binaries as GW sources}\label{sec:boson stars}

\subsection{Equilibrium configurations}
BSs are self-gravitating configurations of a complex scalar field described by the action
\begin{equation}
    \label{action:1}
    S=\int d^4x\sqrt{-g}\left[\frac{R}{16\pi}-{\cal L}_{\rm scalar}\right]\,.
\end{equation}
The shape of the potential in Eq.~\eqref{Lagrangian} determines the properties of the corresponding star. For a simple Klein-Gordon potential $V=\frac{\mu^2}{2}\left|\phi\right|^2$
initially considered in \cite{kaup1968,ruffini1969}, BSs have a maximum mass which scales with the boson mass $m_S\equiv\mu\hbar$ as $M_{\rm max}\sim M_P^2/m_S$. Unless the scalar field is ultralight ($m_S\ll 10^{-11}\,{\rm eV}$), the maximum mass is much smaller than the Chandrasekhar mass for compact astrophysical bodies and, therefore, such objects are often called \emph{mini} BSs. In this work we shall focus on the  case of a quartic potential
\begin{equation}
    \label{potential:2}
    V\left(\left|\phi\right|\right)=\frac{\mu^2}{2}\left|\phi\right|^2+\frac{\lambda}{4}\left|\phi\right|^4\,.
\end{equation}
In the strong-coupling limit $\lambda\gg \mu^2$, the maximum mass of a \emph{static} BS scales as~\cite{colpi1986}
\begin{equation}
    \label{mmax:1}
    M_{\rm max}^{\rm static}\approx0.06\frac{\sqrt{\lambda \hbar}}{m_S^2}M_P^3\approx10^5 M_\odot\sqrt{\lambda \hbar}\left(\frac{{\rm MeV}}{m_S}\right)^2\,.
\end{equation}

Note that, if $\lambda\hbar=1$, varying the boson mass from few percents of a MeV up to a hundred MeV, one can cover the whole spectrum of physical BH masses.\footnote{Following Refs.~\cite{amaro2010,giudice2017}, it is also interesting to note that cosmological observations seem to suggest an interacting dark-matter component, with a cross section per unit mass $0.1 \text{cm}^2/\text{g}\lesssim\sigma/m_{dm}\lesssim 1 \text{cm}^2/\text{g}$, which, for the quartic potential~\eqref{potential:2}, translates to~\cite{giudice2017}
\begin{equation}
    \label{selfDM}
    \lambda\hbar \sim\left(\frac{m_S}{100\,{\rm MeV}}\right)^{3/2}\,. 
\end{equation}
When $m_S\sim 10^3$~eV, this corresponds to supermassive BSs with $M\sim 10^5 M_\odot$ (see Fig.~\ref{fig:paramspace}).
} Hereafter we shall focus only on this regime, which allows for stellar-mass BSs [with mass $M=\mathcal{O}(10M_\odot)$] when $m_S=\mathcal{O}(10^2\,{\rm MeV})$, and for supermassive BSs [with mass $M={\cal O}(10^5 M_\odot$)] when $m_S={\cal O}({\rm MeV})$.

While the boson mass sets the scale of the system, in the strong-coupling limit all properties of the BS depend only on the following combination of the fundamental constants appearing in the action~\eqref{action:1},
\begin{equation}
    \label{MB}
    M_B \equiv \frac{\sqrt{\lambda}}{\mu^2} =\sqrt{\lambda \hbar} \frac{M_P^3}{m_S^2}\,,
\end{equation}
which has the dimension of a mass in our units.

\subsection{GW signatures}\label{sec:signatures}

The structure of a BS differs from that of a BH in several respects, which introduce distinctive features in the gravitational waveforms from coalescing BSs. The main differences can be summarized as follows~\cite{cardoso2019testing}:
\begin{enumerate}[i.]
    \item BSs are less compact than BHs, as measured by their compactness $C=M/R$, where $R$ is the BS effective radius\footnote{At variance with NSs, BSs do not have a hard surface, as the scalar field is spread out
    all over the radial direction. However, it decays exponentially and the configuration is highly localized in a radius $\sim1/\mu$. It is customary to
    define the effective radius $R$ as the radius within which the $99\%$ of the total mass is contained.}. Nonspinning BHs have a compactness $C=M/R=1/2$, while massive BSs have a maximum compactness $C_{\rm max}\approx0.158$ \cite{amaro2010}, which is comparable to the typical compactness of a NS. This also implies that the ``contact" frequency of a binary BS is lower than in the binary BH case. For an equal-mass binary, the GW contact frequency in the point-particle limit reads
    \begin{eqnarray}
        \label{eq:f:contact}
        f_{\rm contact} &=& \frac{1}{2^{3/2}\pi M_t}C^{3/2}\nonumber\\
        &\approx& 1\,{\rm mHz} \left(\frac{10^6M_\odot}{M}\right) \left(\frac{C}{0.15}\right)^{3/2}\,,\label{fcontact}
    \end{eqnarray}
    which for $C\sim0.15$ is similar to the frequency of the innermost stable circular orbit~(ISCO) of a Schwarzschild BH with mass equal to the total mass $M_t=2M$ of the binary, $f_\text{ISCO}=(6^{3/2}\pi M_t)^{-1}$. The latter can be approximately assumed as the transition frequency between the inspiral and the merger in the binary BH case.
    \item The spin $J$ of a BS is quantized\footnote{It has been recently shown that spinning mini BSs made of a scalar field are unstable and decay to their nonspinning state~\cite{sanchis2019}. The instability occurs on dynamical time scales, at least for large compactness. It is unknown whether self-interactions can cure this instability or make it phenomenologically irrelevant.} in units of its Noether charge, the latter existing due to the $U(1)$ symmetry of the Lagrangian~\eqref{Lagrangian}. Nonetheless, in the strong-coupling limit the quantization levels are extremely close to each other, and in practice the dimensionless spin $\chi=J/M^2$ can be treated as a continuous parameter~\cite{ryan1997}.
    \item BSs have a nonvanishing tidal deformability, which expresses the tendency of the object to deform under the action of an external tidal field. In contrast, the tidal deformability --~as measured by the tidal Love numbers~-- of a nonspinning BH is zero\footnote{While this property holds true also for slowly spinning BHs in the axisymmetric case~\cite{Pani:2015hfa,Landry:2015zfa,Pani:2015nua}, it has been recently argued --~using analytical continuation methods~-- that the tidal Love numbers of a spinning BH are nonzero in the nonaxisymmetric case~\cite{LeTiec:2020spy}. Whether and how this affects the gravitational waveform is still an open question. At any rate, even if the tidal deformability of spinning BHs affects the waveform, its (small) fixed value can be used as a baseline for null-hypothesis tests, like the case of the spin-induced quadrupole.}~\cite{Damour:2009vw,binnington2009}.
    \item The higher multipoles of a BH are uniquely determined by its mass $M$ and its dimensionless spin $\chi$. In particular, the quadrupole moment of a Kerr BH is $Q_{\rm Kerr}=-M^3\chi^2$. On the other hand, as we shall review in Sec.~\ref{sec:properties_spin}, the multipole moments of a BS are not necessarily quadratic in the spin and also vary with the compactness.
    \item GWs interact weakly with the scalar field, so that effectively BSs do not absorb gravitational radiation. Therefore, at variance with BHs, tidal heating is absent~\cite{Maselli:2017cmm,Datta:2019epe}.
\end{enumerate}
In brief, the GW signatures of a BS are similar to those of a NS (see Refs.~\cite{Harry:2018hke,Chatziioannou:2020pqz} for some reviews), despite the fact that BSs can be supermassive (and, therefore, are potential exotic sources also for space-based detectors) and, in principle, highly spinning. 

In the next subsections we shall review the above properties more quantitatively and provide useful fits for the quantities that constitute the basic ingredients for the PN waveform model that we shall later use.

\subsubsection{Tidal deformability}
\label{sec:properties_tidal}

The tidal deformability of (static) BSs was computed in \cite{cardoso2017,sennett2017}, which investigated the possibility of using GW measurements of tidal effects to distinguish BSs from BHs and from NSs (see also Ref.~\cite{wade2013}).
The (dimensionful) tidal deformability parameter $\lambda_T$ is defined by~\cite{flanagan2008}
\begin{equation}
    \mathcal{Q}_{ij}=-\lambda_T\mathcal{E}_{ij} \label{Qij}
\end{equation}
where $\mathcal{E}_{ij}$ is the external tidal field and $\mathcal{Q}_{ij}$ is the induced asymptotic quadrupole moment. It is more convenient to work with the dimensionless tidal deformability (hereafter simply tidal deformability) $\Lambda=\lambda_T/M^5$, where $M$ is the mass of the object. A fitting formula for $\Lambda$ in terms of the mass $M$ of the BS and of the parameters $\mu$ and $\lambda$ was obtained in Ref.~\cite{sennett2017}. In the strong coupling limit, the fit reduces to\footnote{Notice that the scalar Lagrangian of \cite{sennett2017} differs from our Eq.~\eqref{action:1} by a factor of 2, which we took into account when rewriting Eq.~\eqref{fit:lambda}.}
\begin{equation}
\label{fit:lambda}
        \frac{M}{M_B}=\frac{\sqrt{2}}{8\sqrt{\pi}}\left[-0.828+\frac{20.99}{\log\Lambda}-\frac{99.1}{\left(\log\Lambda\right)^2}+\frac{149.7}{\left(\log\Lambda\right)^3}\right]\,.
\end{equation}
The above expression can be inverted to find $\Lambda=\Lambda(M_B,M)$. Note that Eq.~\eqref{fit:lambda} has a stationary point for $\Lambda\approx289$, corresponding to $M/M_B\approx0.0611$, which is in agreement with Eq.~\eqref{mmax:1} for the maximum mass. Therefore, the tidal deformability of a massive BS is bounded from below by $\Lambda\gtrsim289$. Notice that $\Lambda$ spans many orders of magnitude as the mass deviates from its maximum value: for example, when $M/M_B=0.02$, $\Lambda\approx1.7\times10^6$. As we show in Appendix~\ref{appendix:1}, this is related to the fact that $\Lambda$ can be expressed in the form $\Lambda\sim k_2/C^5$, where $k_2={\cal O}(0.01-0.1)$ is a numerical factor~\cite{cardoso2017}, and the compactness $C$ can be as low as $0.03$ when $M/M_B=0.02$.

\subsubsection{Spin effects and the quadrupole moment}
\label{sec:properties_spin}

\begin{figure}[hbt!]
    \centering
    \includegraphics[width=0.46\textwidth]{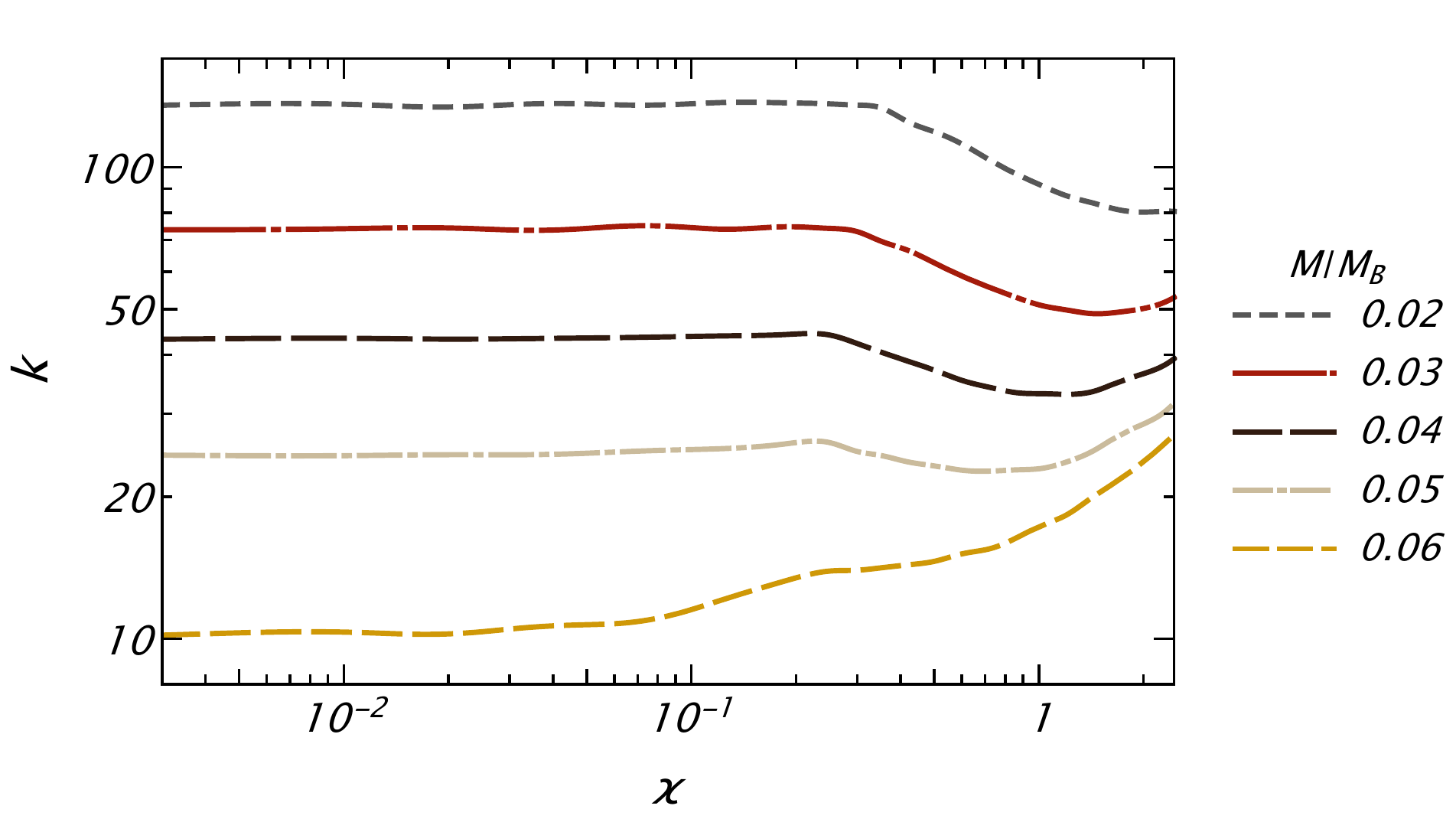}
    \caption{Spin-induced reduced quadrupole moment of a massive BS. The data are extracted and interpolated from Fig.~4 in Ref.~\cite{ryan1997}.}
    \label{fig:ryan:q}
\end{figure}

While $\mathcal{Q}_{ij}$ in Eq.~\eqref{Qij} is a \emph{tidal-induced} quadrupole moment, a spinning self-gravitating body has a \emph{spin-induced} quadrupole moment \cite{laarakkers1999}. An analysis of spinning BSs in the strong-coupling limit was carried out in Ref.~\cite{ryan1997}. The latter contains several important results, which we briefly summarize:
\begin{enumerate}[i.]
    \item The maximum mass of a rotating BS is higher than in the nonspinning case. In Appendix~\ref{appendix:1} we show that $M_{\rm max}$ as a function of the spin is very well approximated by
    \begin{equation}
        \label{mmax:2}
        M_{\rm max}\approx0.06\left(1+0.76\chi^2\right) M_B\,.
    \end{equation}
    \item In the axisymmetric case, the spin-induced quadrupole moment $\mathcal{Q}_{ij}$ can be written in terms of a single scalar quantity $Q$, which for a BS can be parametrized as 
    \begin{equation}
        \label{quadrupole:moment}
        Q = -\kappa(\chi,M/M_B)\chi^2 M^3\,.
    \end{equation}
    At variance with the BH case, the function $\kappa$ is not a constant ($\kappa_{\rm BH}=1$), but it depends on the spin and on the BS mass through the ratio $M/M_B$. The quantity $\kappa$ is shown in Fig.~\ref{fig:ryan:q} as a function of the dimensionless spin $\chi$ for some representative values of $M/M_B$. 
    It is worth noticing that $\kappa$ depends on the parameters of the potential only through the combination $M_B$ defined in~\eqref{MB}, as it is also the case for $\Lambda$. Hereafter we will refer to $\kappa$ as the ``reduced quadrupole moment".
    A similar behavior is exhibited by the spin-induced octupole moment~\cite{ryan1997}. In the following, we will neglect this effect since it affects the GW waveform to higher PN order.
\end{enumerate}
For a BH, $\Lambda=0$ and $\kappa=1$, while for a compact BS $\Lambda\sim (10^2,10^6)$ and $\kappa(\chi\approx0)\sim(10,150)$. Therefore, there is a discontinuity gap between BHs and BSs, and the effects of these parameters on the waveform can be potentially large. Crucially, both $\Lambda$ and $\kappa$ depend only on the object mass (and spin) and on the coupling constant~\eqref{MB}. Therefore, for a given BS model $\Lambda$ and $\kappa$ are not independent quantities.

As shown in Fig.~\ref{fig:ryan:q}, the behavior of $\kappa$ as a function of $\chi$ for fixed values of $M/M_B$ is nonmonotonous at large spins, whereas the behavior of $\kappa_0\equiv \kappa(\chi\approx0)$ as a function of $M/M_B$ is simpler. As shown in Appendix~\ref{appendix:1}, after mapping the dependence of $M/M_B$ to $\Lambda$ one finds a linear fit for the logarithmic quantities,
\begin{equation} 
    \label{love:q:1}
    \log\kappa_0 \approx 0.61+0.3\left(\log\Lambda\right)\,.
\end{equation}
In practice, this fit relates the reduced quadrupole moment of a slowly-spinning massive BS to its Love number. The full quadrupole moment is corrected by ${\cal O}(\chi^4)$ terms.
It is interesting that the $(\log\kappa_0)$-$(\log\Lambda)$ relation is approximately linear. This is reminiscent of the case of NSs, for which the ``Love-Q" relations have a similar form~\cite{yagi2013,yagi2013_2}. In that case, the coefficients of the fit are nearly universal for different equations of state of the star. In a future work, we will explore this issue for BSs with different scalar potentials~\cite{massimo}.

In the following, we will make use of Eqs.~\eqref{love:q:1} and \eqref{fit:lambda} to express all finite-size effects in the waveform in terms of the single parameter $M_B$. 
In the next section, we review how these effects (as well as the tidal heating) enter in the PN approximation of the GW waveform.

\section{PN corrections to the waveform}
\label{sec:waveform}
We focus on the inspiral phase of the signal emitted by a compact binary coalescence, adopting the PN expanded TaylorF template in the frequency domain~\cite{Damour:2000zb,Damour:2002kr,Arun:2004hn}
\begin{equation}
    \label{waveform:1}
    h(f)=\mathcal{A}(f)e^{i\psi(f)}\,.
\end{equation}
The amplitude $\mathcal{A}$ and the phase $\psi$ are expanded as power series 
in the orbital velocity $v=\left(\pi M_t f\right)^{1/3}$, where $M_t=M_1+M_2$ is the total mass of the binary and $M_i$ is the mass of the $i$th binary component (we assume $M_1\geq M_2$). A term proportional to $v^n$ corresponds to the $n/2$-PN order of the approximation. 
In our analysis we retain only the dominant (Newtonian) term in amplitude, also 
averaging over the orientation and the polarization angles, which specify the source's 
position with respect to the detector, such that
\begin{equation}
    \label{rms:amplitude}
    \mathcal{A}=\frac{M_t^2}{D_L}\sqrt{\frac{\pi\eta}{30}}\left(\pi M_tf\right)^{-7/6}\ ,
\end{equation}
where $D_L$ is the luminosity distance and $\eta=M_1 M_2/M_t^2$ is the symmetric 
mass ratio. For LISA, Eq.~\eqref{rms:amplitude} must also be multiplied by an 
additional factor of $\sqrt{3}/2$ in order to account for the triangular geometry of the detector~\cite{Berti:2004bd}.

The BS signatures described in Sec.~\ref{sec:signatures} 
affect the signal's phase at different PN orders, reflecting the frequency content of each effect. At small frequencies, where lower PN terms play a more significant role, sources behave as point particles and the details on their internal structure are effaced~\cite{damourEffac,blanchet2014review}. For larger frequencies, however, finite-size terms induced by spin-quadrupole and tidal effects become relevant.  
Changes in the waveform due to the BS structure add linearly to the BH 
phase, namely:
\begin{equation}
\psi(f)= \psi_{\textnormal{BH}}(f)+\psi_{\kappa}(f)+\psi_{\Lambda}(f)\ ,
\end{equation}
where $\psi_\kappa$ and $\psi_{\Lambda}$ identify the 
modifications induced by the spin-quadrupole and by the tidal terms, respectively. 
In our analysis, we consider a 3.5PN expanded phase $\psi_\textnormal{BH}$~\cite{mishra2016}, which includes spin-orbit, spin-spin (up to 3PN), and cubic spin 
corrections~\cite{Khan:2015jqa,Isoyama:2017tbp}. 
Furthermore, since BSs do not absorb GWs efficiently, for consistency in the ``BH part'' of the waveform ($\psi_{\textnormal{BH}}$), we ignore the tidal-heating term which enters the phase with a $v^5\log v$ (respectively, $v^8\log v$) correction for a spinning (nonspinning) BH.

The dominant tidal correction enters the waveform at 5PN order, hence it is 
suppressed by a factor $v^3$ with respect to the 3.5PN phase $\psi_\textnormal{BH}$. 
However, the potentially large values
of $\Lambda$ render tidal effects comparable with the rest of the point-particle expansion. 
The leading tidal correction to the phase is given by~\cite{flanagan2008,favata2014}
\begin{equation}
    \label{tidal:1}
    \psi_{\Lambda}=-\frac{117}{256\eta}\tilde{\Lambda}\left(\pi M_t f\right)^{5/3}
\end{equation}
where
\begin{equation}
    \label{tidal:2}
    \tilde{\Lambda}=\frac{16}{13}\left[\left(1+\frac{12}{q}\right)\frac{M_1^5}{M_t^5}\Lambda_1+\left(1+12q\right)\frac{M_2^5}{M_t^5}\Lambda_2\right]
\end{equation}
is an effective total tidal deformability, and $q=M_1/M_2\geq1$ is the binary mass ratio; the normalization of $\tilde{\Lambda}$ is chosen so that $\tilde{\Lambda}=\Lambda_1=\Lambda_2$ for an equal mass binary. Indeed, note that $\Lambda_1$ and $\Lambda_2$ are not independent, since $\Lambda_i=\Lambda_i(M_i, M_B)$, so for a given theory they are fixed once the masses are known.\footnote{The same situation occurs for NSs once the equation of state is fixed. We remind that in the BS case the role of the equation of state is played by the potential $V(|\phi|)$.} 
The next-to-leading tidal correction enters at 6PN order and depends on both $\tilde{\Lambda}$ and on a second combination
$\delta\tilde{\Lambda}$ of $\Lambda_1$ and $\Lambda_2$ \cite{Lackey:2014fwa}, and also on the magnetic tidal Love numbers~\cite{Yagi:2013sva,pani2018magnetic,abdelsalhin2018,forteza2018impact,Poisson:2020mdi}. The correction $\delta\tilde{\Lambda}$ has been, in general, neglected within previous studies on BS waveforms since it is subdominant 
in the PN expansion, and if $\Lambda_1$ and $\Lambda_2$ are treated as independent quantities, it would introduce an extra waveform parameter . However, since in our model $\Lambda_i=\Lambda_i(M_B,M_i)$, including 6PN corrections does not introduce any extra parameter in the waveform, and at the same may actually lead to an overall improvement on the constraints that characterize the BS's structure. On the other hand, the effect of the magnetic tidal deformability is typically negligible since the magnetic tidal Love numbers are much smaller than the standard (electric) ones~\cite{forteza2018impact,cardoso2017}. For this reason, we shall neglect the magnetic-tidal contribution to the 6PN waveform.

The spin-induced quadrupole moments affect 
the GW phase already at 2PN order~\cite{wade2013,krishnendu2017,Kastha:2018bcr,Raposo:2020yjy}. 
For aligned spin binaries, the dominant quadrupole contribution reads
\begin{equation}
\label{phase}
 \psi_{\rm\kappa} = -\frac{75}{64} \frac{\left(\kappa_1M_1^2\chi_1^2+\kappa_2M_2^2\chi_2^2\right)}{M_1 M_2} \left(\pi M_t f\right)^{-1/3}\,,
\end{equation}
where $\kappa_i$ is the spin-induced reduced quadrupole of the $i$th body (the binary BH case corresponds to $\kappa_1=\kappa_2=1).$
We incorporate also the first subdominant correction (appearing at 3PN) as given in Ref.~\cite{krishnendu2017}. 
We neglect the contributions from the spin-induced octupolar corrections, which are subleading relative to the quadrupolar corrections.  

Finally, we restrict our analysis to nonprecessing binaries, i.e., we assume that 
the individual spins are aligned with the binary orbital angular momentum.

\section{Measurability of the binary parameters}
\label{sec:fisher}
In this section we briefly review the basic properties of 
the Fisher information matrix formalism that we use to infer the uncertainties 
on the waveform parameters \cite{cutler1994,vallisneri2008}, 
and we summarize the results of previous applications 
of such formalism to the problem of distinguishing 
BS binaries from their BH counterparts.

We assume that a detection criterion for a GW signal 
$h(t,\vec{\theta})$ has been met, providing us with the 
best estimates for the source parameters $\vec{\theta}_0$ (as masses, spins, distance, orientation angles, etc.). 
Let $h(f,\vec{\theta}_0)$ be the waveform in the 
frequency domain evaluated at $\vec{\theta}=\vec{\theta}_0$.
The signal-to-noise ratio (SNR) associated to the 
detection is given by

\begin{equation}
    \label{snr:1}
    \left(\text{SNR}\right)^2=\left(h\left|\,h\right.\right)\ ,
\end{equation}
where
\begin{equation}
    \label{inner:product}
    \left(h_1\left|\,h_2\right.\right)=4\Re\int_{f_{\rm min}}^{f_{\rm max}} df\frac{h_1(f)^\star h_2(f)}{S_n(f)}
\end{equation}
is the waveform inner product defined over the detector's 
noise spectral density $S_n(f)$. 
For LISA, we choose the range of integration in the frequency domain as follows:
\begin{subequations}
\begin{align}
    & f_{\rm min} = \text{\rm max}\left(10^{-4}\,\text{Hz},f_{\rm obs}\right)\\
    & f_{\rm max} = \text{min}\left(1\,\text{Hz}, f_\text{ISCO}\right)\ ,
\end{align}
\end{subequations}
where we remind $f_\text{ISCO}=(6^{3/2}M_t\pi)^{-1}$, whereas 
$f_{\rm obs}$ is determined by requiring that the observation lasted 
$T_{\rm obs}=1$ year before the binary reached the ISCO frequency, namely (using a Newtonian approximation~\cite{berti2005})
\begin{equation}
    \label{fobs}
    f_{\rm obs}=4.149\times10^{-5}\left(\frac{\mathcal{M}}{10^6 M_\odot}\right)^{-5/8}\left(\frac{T_{\rm obs}}{1\,\text{yr}}\right)^{-3/8}\,\text{Hz}\,.
\end{equation}

On the other hand, for the analysis with the ET we use
\begin{equation}
    f_{\rm min} = 3\,\text{Hz}\quad \ ,\quad
    f_{\rm max} = f_\text{ISCO}\,.
\end{equation}

In the limit of large SNR, the best estimates $\vec{\theta}_0$ 
are unbiased, meaning that they approach the true values. 
If we also assume that the instrumental noise is Gaussian,\footnote{Technically speaking, one also has to assume that the priors over $\theta^\mu$ are flat, which is approximately valid if the scale over which the priors change is smaller than the scale over which \eqref{p:theta:1} changes.} then the posterior distribution of the waveform parameters 
$\vec{\theta}$ is described by a normal distribution peaked 
around $\vec{\theta}_0$ \cite{vallisneri2008}:
\begin{equation}
    \label{p:theta:1}
    p\left(\theta\left|\mathbf{s}\right.\right)\propto e^{-\frac{1}{2}(\theta-\theta_0)^\mu\Gamma_{\mu\nu}(\theta-\theta_0)^\nu}
\end{equation}
where $\bf{\Gamma}$ is the Fisher information matrix defined 
by
\begin{equation}
    \label{fisher:1}
    \Gamma_{\mu\nu}=\left(\frac{\partial h}{\partial\theta^\mu}\left|\,\frac{\partial h}{\partial\theta^\nu}\right.\right)\,.
\end{equation}
The 1-$\sigma$ uncertainty on $\Delta\vec{\theta}=\vec{\theta}-\vec{\theta}_0$ is given by
\begin{equation}
    \label{sigma:1}
    \sigma_j\equiv\sqrt{\left<(\Delta\theta^j)^2\right>}=\sqrt{\left(\bf{\Gamma}^{-1}\right)_{jj}}\,.
\end{equation}

The result~\eqref{sigma:1} is formally valid only in the limit of infinite SNR,
while for large SNR it is understood to provide only an order-of-magnitude estimate~\cite{vallisneri2008}. A rigorous approach would require a Bayesian analysis (see, e.g., \cite{rodriguez2013}). However, Bayesian simulations are computationally costly, while the Fisher formalism is extremely cheap; therefore, the latter is useful to understand if the problem is promising enough to be addressed with a more rigorous Bayesian approach.

Previous studies focused on waveforms' changes 
induced by tidal interactions \cite{cardoso2017,sennett2017,Pani:2019cyc}, 
horizon absorption effects \cite{Maselli:2017cmm,Datta:2019epe}, and 
spin-induced multipole moments \cite{krishnendu2017,Kastha:2018bcr,krishnendu2019,krishnendu2019_2,Kastha:2019brk}.
In particular, previous work considered only a single effect at the time, focusing for instance on the detectability of $\tilde\Lambda$ or $\kappa$. However, given a BS model, both $\Lambda$ and $\kappa$ (as well as all other inspiral waveform parameters) depend only on the masses and spins of the binary. Therefore, including all effects at once does not increase the number of waveform parameters, on the contrary, it might help breaking degeneracy and improving the accuracy of the template.

\begin{figure*}[hbt!]
    \centering
    \includegraphics[width=0.99\textwidth]{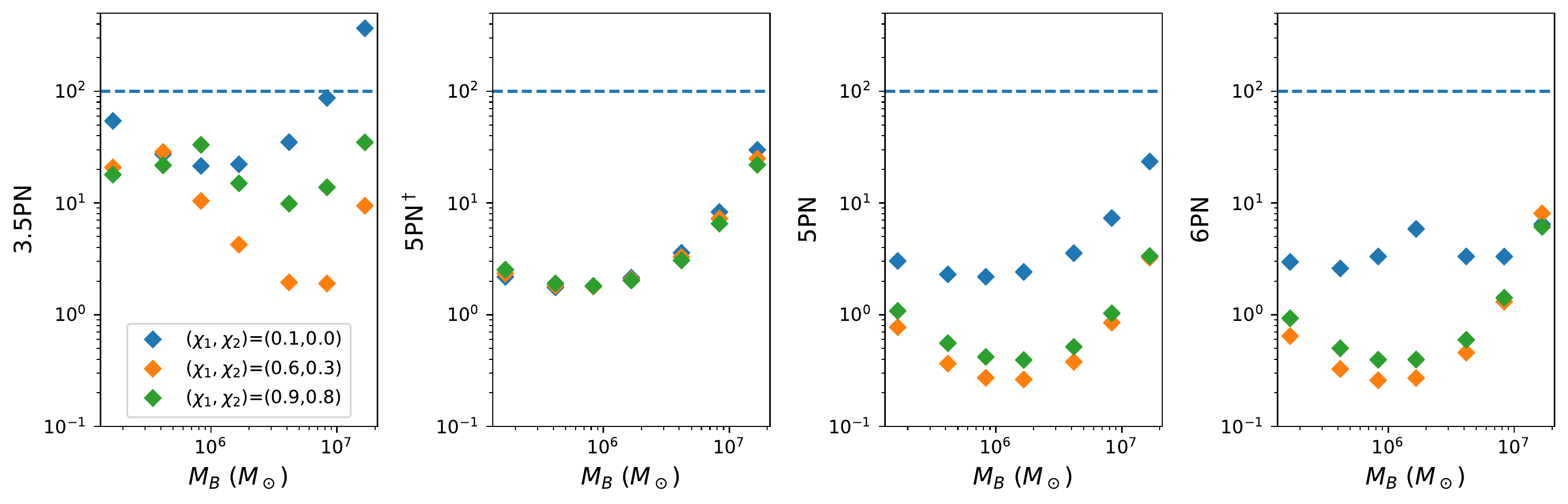}
    \caption{Relative percentage errors $\sigma_{M_B}/M_B[\%]$ for binary BSs with a coupling parameter in the range $M_B\in1.7\times[10^5,10^7]M_\odot$, located at a luminosity distance $D_L=1$~Gpc and observed with LISA. From left to right, panels refer to different terms included in the PN waveform model, see text for details.}
    \label{fig:results:1}
\end{figure*}
\begin{figure*}[hbt!]
    \centering
    \includegraphics[width=0.99\textwidth]{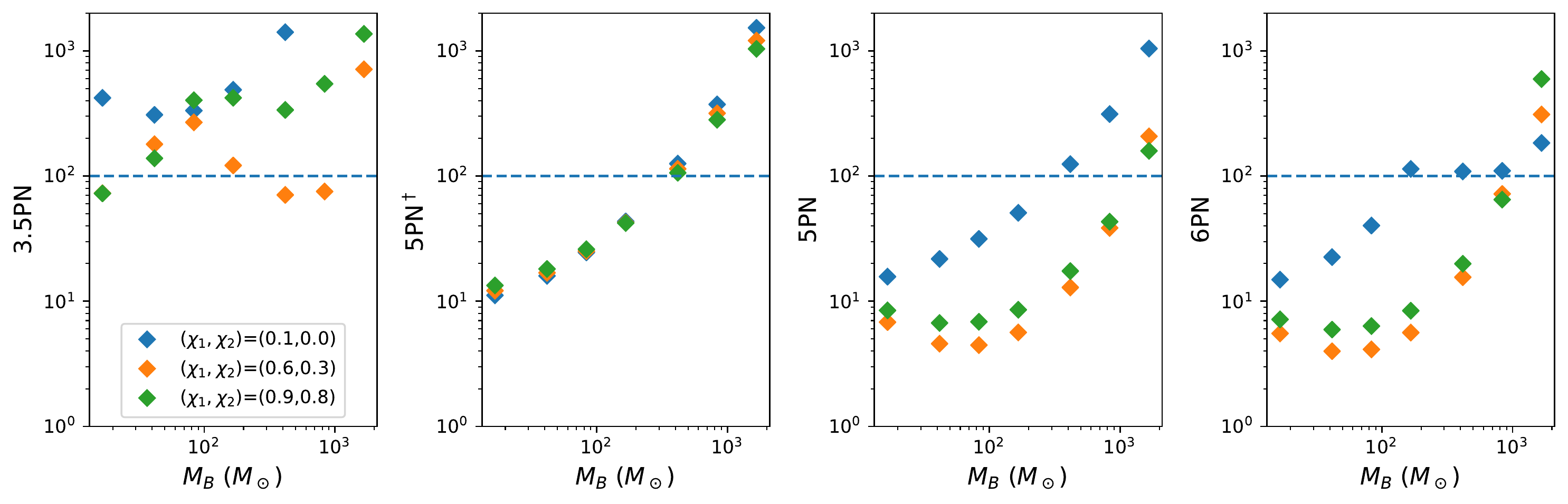}
    \caption{Same as in Fig.~\ref{fig:results:1} but for binary BSs with coupling parameter in the range $M_B\in1.7\times[10,10^3]M_\odot$, located at a luminosity distance $D_L=500$~Mpc and observed with the ET.}
    \label{fig:results:2}
\end{figure*}

\section{Results}
\label{sec:results}
Here we present the results of our Fisher-matrix analysis aimed 
to assess the ability of measuring the fundamental coupling 
constants of a scalar field theory using GW detections of binary BSs by LISA and the ET. 
For concreteness, we restrict our attention to slightly unequal-mass 
binaries, such that $M_1\approx0.06 M_B$ is approximately the maximum 
mass for nonspinning BSs\footnote{From Eq.~\eqref{fit:lambda}, $\Lambda$ has an extremum at $M=M_{\rm max}^{\rm static}$, so its derivative with respect to $M/M_B$ 
diverges at that point, thus rendering the Fisher matrix ill-defined. 
To overcome this problem, we choose $(M_{\rm max}^{\rm static}-M_1)/M_{\rm max}^{\rm static}=10^{-4}$.} and $M_2=0.05 M_B$, 
with the mass ratio equal to $q=1.2$. We proceed by increasing levels of 
complexity:
\begin{enumerate}[i.]
    \item First, we consider the case in which the waveform has corrections only from the spin quadrupole moments, ignoring tidal deformabilities. We refer to this case as the ``3.5PN test" because the waveform is expanded only up to 3.5PN order.
    \item Then, we consider the case in which the waveform has corrections only from the tidal deformabilities, ignoring spin-quadrupole moment terms. We refer to this case as the ``5PN$^\dagger$ test" because we include only the leading 5PN tidal correction, and the $\dagger$ refers to the absence of spin-quadrupole corrections.
    \item In the third case, we incorporate both quadrupole and tidal corrections, the latter only at the leading 5PN order. We refer to this case as the ``5PN test".
    \item Finally, we also include the 6PN tidal correction (though we neglect the 6PN contribution from the smaller magnetic tidal deformabilities). We refer to this case as the ``6PN test".
\end{enumerate}
For all cases, the independent parameters of the Fisher matrix are
\begin{equation}
    \vec{\theta}=\left(\mathcal{A},t_c,\phi_c,\log\mathcal{M},\log\eta,\chi_s,\chi_a,M_B\right)\,,
\end{equation}
where $t_c$ and $\phi_c$ are the coalescence time and phase, $\mathcal{M}=(M_1M_2)^{3/5}/M_t^{1/5}$ is the chirp mass, and $\chi_{s,a}=(\chi_1\pm\chi_2)/2$ are the symmetric and antisymmetric spin components. By considering the amplitude $\mathcal{A}$ as an independent parameter, 
it decouples from the other variables \cite{cutler1994}, 
such that the Fisher matrix becomes block-diagonal 
$\Gamma_{i{\cal A}}=\delta_{i{\cal A}}$;
therefore we will not report the uncertainties for $\mathcal{A}$.

For each choice of $M_B$, we use Eq.~\eqref{fit:lambda} to compute the individual 
tidal deformabilities, and Eq.~\eqref{love:q:1} to compute the corresponding 
reduced quadrupole moments. This is not entirely consistent because 
Eq.~\eqref{love:q:1} is valid only at small spins, while for generic spins 
$\kappa$ has the complicate dependence shown in Fig.~\ref{fig:ryan:q}. 
However, it is difficult to extract the dependence on $M_B$ from Fig.~\ref{fig:ryan:q}, 
especially due to the fact that data are not dense enough. Work in this direction is in progress and will be reported elsewhere~\cite{massimo}. Here, for simplicity, we assumed 
$\kappa\approx\kappa_0$, which amounts to neglect ${\cal O}\left(\chi^4\right)$ 
corrections in the expression \eqref{quadrupole:moment} for the quadrupole moment. This approximation is valid for moderately small spins. 

For LISA, we consider a coupling in the range $M_B\in1.7\times[10^5,10^7] M_\odot$, which corresponds to a maximum mass for nonspinning BSs $M_{\rm max}^{\rm static}\in[10^4,10^6]M_\odot$ at a fixed luminosity distance $D_L=1$Gpc. For the ET we consider $M_{\rm max}^{\rm static}\in[1,100]M_\odot$ at a fixed luminosity distance $D_L=500$Mpc. In both cases, we consider three different spin combinations $(\chi_1,\chi_2)=(0.1,0),(0.6,0.3),(0.9,0.8)$. Since we consider relatively small distances, the effect of the cosmological redshift is small, therefore, in the Fisher matrix we will use the values of the masses in the source frame.

Figures~\ref{fig:results:1} and \ref{fig:results:2} show the relative uncertainties on $M_B$ in the entire mass ranges considered, respectively, for LISA and the ET. We can understand the importance of the various contributions to the waveform as follows. First, tidal corrections have a larger impact than spin-quadrupole ones on the accuracy in $\sigma_{M_B}$: this can be seen by comparing the 3.5PN and 5PN$^\dagger$ panels, from which it is evident that for the 5PN$^\dagger$ model the uncertainties are smaller by an order of magnitude (and even more for small spins). The results of the 5PN$^\dagger$ model are in agreement with those of Ref.~\cite{cardoso2017}, as expected. In particular, we observe that they vary only mildly with the spin magnitude, as it is expected from the fact that the 5PN tidal correction to the phase does not depend explicitly on the spins.
Second, including the spin-induced quadrupole moment helps to break the degeneracy with the spin among different PN terms in the waveform, leading to an improvement in the overall measurability at moderate and high spins: indeed, by contrasting the full 5PN results with the reduced 5PN$^\dagger$, we can observe that the relative uncertainty in $M_B$ improves by an order of magnitude when $(\chi_1,\chi_2)=(0.6,0.3),(0.9,0.8)$. Finally, the inclusion of 6PN tidal corrections leads to a moderate variation around $40\%$ on $\sigma_{M_B}$, showing that the values inferred from the 5PN case are robust against the inclusion of the next-order tidal correction. This also confirms that the PN series is converging well also at $f\approx f_{\rm ISCO}$, suggesting that a PN waveform approximant is sufficiently accurate up to those frequencies.

Crucially, when all corrections are included, the uncertainties on $M_B$ are at subpercent and percent level, for LISA and the ET respectively, in the most optimistic configurations and exceed the value of $100\%$ only for the highest masses in the case of ET.
This means that, in the considered mass range, a putative binary BS detection can be used to measure the coupling $M_B=\sqrt{\lambda}/\mu^2$ with great precision. We remind that the errors coming from the Fisher analysis scales linearly with the luminosity distance, so assuming (say) a distance $D_L=10$~Gpc for LISA binaries would decrease the SNR and increase the errors by a factor of $10$.
Given the small errors in Figs.~\ref{fig:results:1} and \ref{fig:results:2}, even in the more conservative case in which the event is at ${\cal O}(10)\,{\rm Gpc}$ (for LISA) or at ${\cal O}(1)\,{\rm Gpc}$ (for the ET), the measurement of $M_B$ would still be accurate at ${\cal O}(10)\%$ level or better.

The dependence of the constraints on $M_B$ can be understood as follows. The detectability of a signal will be enhanced when the corresponding contact frequency falls within the most sensitive band of the corresponding detector. Using Eq.~\eqref{eq:f:contact} for $f_{\rm contact}$ and approximating $M\approx 2M_{\rm max}^{\rm static}\approx0.12M_B$, we obtain an order of magnitude estimate for the most sensitive regions of $M_B$. For example, the LISA detector will be most sensitive in the frequency band $[10^{-3},10^{-1}]$ Hz, corresponding to $M_B\approx[10^5,10^7] M_\odot$. On the other hand, the ET will be mostly sensitive in the band $[50,500]$ Hz, giving $M_B\approx[20,200] M_\odot$. This reasoning exemplifies why different detectors can probe different parameter regions, while also explaining the local minima in constraints of Figs.~\ref{fig:results:1} and \ref{fig:results:2}.

In Table~\ref{tab:fiducial} we also show the results of the Fisher analysis for LISA for a fiducial binary BS system with $M_B\approx8.3\times10^6 M_\odot$ ($M_{\rm max}^{\rm static}=5\times10^5 M_\odot$), corresponding to ${\rm SNR}\approx1.5\times 10^4$. Similarly, Table~\ref{tab:fiducial:2} shows the results for the ET, for a fiducial system with $M_B\approx8.3\times10^{2}M_\odot (M_{\rm max}^{\rm static}=50M_\odot$) and corresponding ${\rm SNR}\approx620$. 

We have also checked whether, in the mass range considered for the ET, a successful measurement can be achieved with the advanced LIGO detector at design sensitivity. However, we find that the relative uncertainties in $M_B$, as measured with the aLIGO, always lie above $100\%$ with the exception of some high spins configurations which are only marginally detectable.
This is not only due to the globally lower SNR relative to ET, but more importantly to LIGO's low-frequency cutoff, which is higher than the ET's. Higher sensitivity at low frequency helps significantly to measure the low-PN parameters, and in turn to improve the measurement of high-PN parameters at higher frequency. We confirmed this expectation by artificially truncating the ET noise curve at higher low-frequency cutoff, finding that the constraints on $M_B$ quickly deteriorate as the low-frequency cutoff increases.
To summarize, our analysis suggests that future-generation GW observatories are necessary to make precise measurements of the fundamental couplings of a BS model.

\begin{table*}[p!]
    \begin{ruledtabular}
        \begin{tabular}{ccccccccc}
         Test &$(\chi_1,\chi_2)$ &$\sigma_{t_c}$[sec] &$\sigma_{\phi_c}$[rad] &$\sigma_{\log\mathcal{M}}[\%]$ &$\sigma_{\log\eta}[\%]$ &$\sigma_{\chi_s}$ &$\sigma_{\chi_a}$ &$\sigma_{M_B}/M_B[\%]$\\
         \hline
         3.5PN &$(0.1,0)$ &$18.8$ &$0.302$ &$1.41\times10^{-2}$ &$8.90$ &$0.273$ &$2.42$ &$86.5$\\
         5PN$^\dagger$ &$(0.1,0)$ &$28.8$ &$0.982$ &$6.57\times10^{-3}$ &$2.46$ &$9.66\times10^{-3}$ &$0.141$ &$8.27$\\
         5PN &$(0.1,0)$ &$31.1$ &$1.05$ &$6.18\times10^{-3}$ &$2.16$ &$2.00\times10^{-2}$ &$0.254$ &$7.27$\\
         6PN &$(0.1,0)$ &$28.2$ &$1.37$ &$1.23\times10^{-3}$ &$0.952$ &$7.68\times10^{-3}$ &$7.08\times10^{-2}$ &$3.23$\\
         \hline
         3.5PN &$(0.6,0.3)$ &$30.5$ &$4.78$ &$5.31\times10^{-3}$ &$0.378$ &$0.253$ &$2.23$ &$1.90$\\
         5PN$^\dagger$ &$(0.6,0.3)$ &$33.4$ &$1.09$ &$6.14\times10^{-3}$ &$2.15$ &$1.02\times10^{-2}$ &$0.202$ &$7.23$\\
         5PN &$(0.6,0.3)$ &$27.9$ &$4.54$ &$5.30\times10^{-3}$ &$0.246$ &$0.245$ &$2.16$ &$0.829$\\
         6PN &$(0.6,0.3)$ &$35.4$ &$6.98$ &$6.86\times10^{-3}$ &$0.376$ &$0.368$ &$3.26$ &$1.28$\\
         \hline
         3.5PN &$(0.9,0.8)$ &$33.0$ &$5.49$ &$3.14\times10^{-3}$ &$2.28$ &$0.137$ &$1.01$ &$13.8$\\
         5PN$^\dagger$ &$(0.9,0.8)$ &$35.6$ &$1.13$ &$5.89\times10^{-3}$ &$1.94$ &$1.84\times10^{-2}$ &$0.210$ &$6.52$\\
         5PN &$(0.9,0.8)$ &$15.2$ &$0.985$ &$4.57\times10^{-3}$ &$0.297$ &$3.84\times10^{-2}$ &$0.287$ &$1.01$\\
         6PN &$(0.9,0.8)$ &$25.2$ &$1.41$ &$5.43\times10^{-3}$ &$0.407$ &$5.14\times10^{-2}$ &$0.390$ &$1.39$\\
    \end{tabular}
    \caption{Fisher matrix uncertainties for a binary BS with fundamental couplings such that $0.06M_B=5\times10^5 M_\odot$, individual masses $M_1/M_B\approx0.06$ and $M_2/M_B=0.05$, located at a luminosity distance $D_L=1$~Gpc and observed with LISA. The corresponding SNR is $\approx1.5\times 10^4$.}
    \label{tab:fiducial}
    \end{ruledtabular}
\end{table*}

\begin{table*}[p!]
    \begin{ruledtabular}
        \begin{tabular}{ccccccccc}
         Test &$(\chi_1,\chi_2)$ &$\sigma_{t_c}$[sec] &$\sigma_{\phi_c}$[rad] &$\sigma_{\log\mathcal{M}}[\%]$ &$\sigma_{\log\eta}[\%]$ &$\sigma_{\chi_s}$ &$\sigma_{\chi_a}$ &$\sigma_{M_B}/M_B[\%]$\\
         \hline
         3.5PN &$(0.1,0)$ &$8.71\times10^{-2}$ &$14.9$ &$0.914$ &$495$ &$14.7$ &$130$ &$4.74\times10^3$\\
         5PN$^\dagger$ &$(0.1,0)$ &$0.122$ &$45.8$ &$0.353$ &$111$ &$0.237$ &$4.45$ &$373$\\
         5PN &$(0.1,0)$ &$0.132$ &$48.4$ &$0.324$ &$92.1$ &$0.849$ &$11.0$ &$310$\\
         6PN &$(0.1,0)$ &$9.22\times10^{-2}$ &$4.17$ &$4.36\times10^{-2}$ &$31.7$ &$0.258$ &$2.39$ &$107$\\
         \hline
         3.5PN &$(0.6,0.3)$ &$0.124$ &$197$ &$0.313$ &$16.6$ &$10.9$ &$96.0$ &$75.4$\\
         5PN$^\dagger$ &$(0.6,0.3)$ &$0.140$ &$49.3$ &$0.324$ &$94.1$ &$0.528$ &$9.89$ &$317$\\
         5PN &$(0.6,0.3)$ &$0.128$ &$201$ &$0.313$ &$11.1$ &$11.1$ &$97.7$ &$37.7$\\
         6PN &$(0.6,0.3)$ &$0.197$ &$376$ &$0.487$ &$20.7$ &$20.2$ &$179$ &$70.4$\\
         \hline
         3.5PN &$(0.9,0.8)$ &$0.130$ &$218$ &$0.195$ &$90.1$ &$5.63$ &$41.3$ &$548$\\
         5PN$^\dagger$ &$(0.9,0.8)$ &$0.148$ &$50.5$ &$0.309$ &$83.5$ &$0.899$ &$9.90$ &$281$\\
         5PN &$(0.9,0.8)$ &$6.23\times10^{-2}$ &$37.7$ &$0.250$ &$12.4$ &$1.63$ &$11.9$ &$42.3$\\
         6PN &$(0.9,0.8)$ &$0.113$ &$60.0$ &$0.322$ &$18.6$ &$2.38$ &$17.8$ &$63.5$\\
    \end{tabular}
    \caption{Same as Table~\ref{tab:fiducial} but for a binary BS with fundamental couplings such that $0.06M_B=50 M_\odot$, located at a luminosity distance $D_L=500$~Mpc and observed with the ET. In this case the corresponding SNR is $\approx620$.}
    \label{tab:fiducial:2}
    \end{ruledtabular}
\end{table*}

\section{Concluding remarks}
\label{sec:conclusion}
We have estimated the ability of measuring the fundamental coupling constants of a scalar field theory through GW measurements of binary BSs. If such binaries exist and merge, future detectors such as LISA and the ET have the potential to measure the properties of the component BSs with great precision. Such measurements can be mapped into constraints on the fundamental coupling constants of the scalar field theory. As depicted in Fig.~\ref{fig:paramspace}, stellar-mass and supermassive BSs probe different regions of the parameter space, so future LISA and ET measurements will be complementary.

Our estimates are obtained through a Fisher matrix analysis, which is only an approximate treatment (although, given that LISA will reach ${\rm SNR}\gg100$, the approximation should be less dramatic than in the case of, e.g., aLIGO). A more accurate parameter estimation would require using Bayesian techniques. The latter can be also used to compute Bayes factors and perform model selection between different hypotheses for the nature of the binary: BS-BS, BH-BH, BH-BS, or to perform model selection among different BS models. Another source of approximation is that we excluded the merger-ringdown part of the spectrum from our waveform model, and only included the PN corrections to the inspiral. Therefore, in the future, it would be important to incorporate phenomenological completions of the waveform to the merger-inspiral region, possibly informed by numerical simulations (see, e.g., Refs.~\cite{palenzuela2008,palenzuela2017,sanchis2019collisions}). The merger of two BSs shows some distinctive features compared to its BH counterpart, which can be used to confidently determine the nature of the coalescing binaries, and thus subsequently focus on a precision analysis of the inspiral signal to extract information about the model parameters.

Indeed, we have focused on massive BSs, but a natural extension of our work is to consider different classes of the scalar potential $V$, for example mini BSs~\cite{kaup1968,ruffini1969}, solitonic BSs~\cite{friedberg1987}, or the recently studied model of axion BSs~\cite{guerra2019axion,delgado2020rotating}. The latter two models allow for very compact BSs, which are expected to have properties closer to those of a BH.
When spinning, mini BSs are unstable and decay to their ground nonspinning case, at least for certain compactnesses~\cite{sanchis2019}. Strong self-interactions might cure this instability or make it phenomenologically irrelevant.

Likewise, BSs exist also for other integer-spin fundamental fields, such as Proca~\cite{brito2016proca,herdeiro2020tidal} and massive spin-$2$~\cite{Aoki:2017ixz} fields. Interestingly enough, at variance with their scalar counterpart, spinning Proca stars are not unstable, due to the spherical topology of the Proca field profile in the ground state as opposed to the toroidal topology of a scalar field~\cite{sanchis2019}. Our methods can be directly applied also to Proca stars or more generic BSs, provided a theoretical estimate of the tidal deformability and spin quadrupole is available for these models.

We also highlighted the existence of a simple phenomenological relation between the tidal deformability and the spin-induced quadrupole moment of static BSs described by the quartic potential~\eqref{potential:2}. The relation is a straight line in log-space, reminiscent of the approximately universal Love-Q relations for NSs~\cite{yagi2013,yagi2013_2}. We believe that such a remarkably simple behavior deserves further investigation. A computation of quadrupole moments for spinning mini BSs, solitonic BSs, or axion BSs would definitely reveal if such a relation is approximately universal also in the BS case, or it is just an accident of the potential~\eqref{potential:2}.

\begin{acknowledgments}
We thank Monica Colpi for the interesting discussion.
We acknowledge financial support provided under the European Union's H2020 ERC, Starting Grant Agreement No.~DarkGRA--757480, under the MIUR PRIN and FARE programmes (GW-NEXT, CUP:~B84I20000100001), and support from the Amaldi Research Center funded by the MIUR program `Dipartimento di Eccellenza" (CUP:~B81I18001170001).
\end{acknowledgments}

\appendix
\section{Properties of BSS}
\label{appendix:1}
In this appendix we summarize our fits for various BS quantities, obtained by collecting and analyzing the data already existing in the literature. In particular, we will refer to the plots\footnote{The data in \cite{ryan1997} are not tabulated, therefore we extracted them directly from the figures using a coordinate-location software.} in Ref.~\cite{ryan1997} for the properties of spinning BSs, and to the fits given in Ref.~\cite{sennett2017} for the of tidal Love numbers. We will show that several interesting relations emerge.
\subsection{Maximum mass} 
\begin{figure}[hbt!]
    \centering
    \includegraphics[width=0.4\textwidth]{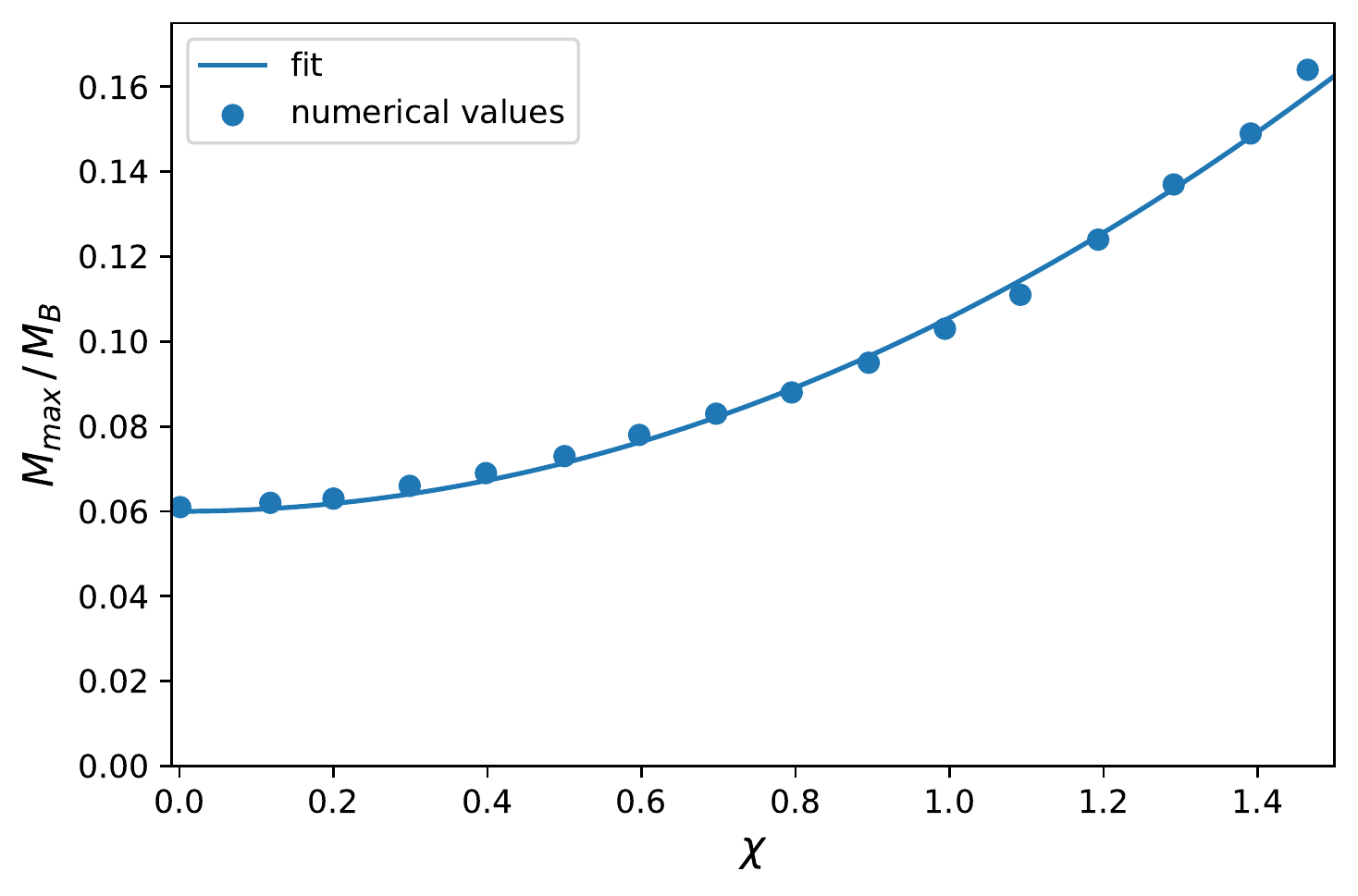}
    \caption{Maximum BS mass $M_{\rm max}/M_B$ as a function of the spin $\chi$ as given by the numerical data (blue spots) and by the parabolic fitting formula \eqref{mmax:2} [continuous line]. The numerical data were extracted from Fig.~1 in Ref.~\cite{ryan1997}.}
    \label{fig:mmax}
\end{figure}

The dependence of the maximum mass on the spin was studied in \cite{ryan1997}. As shown in Fig.~\ref{fig:mmax}, we found that the numerical data are well fitted by a parabolic formula in $\chi$,
\begin{equation}
    M_{\rm max}=0.06\left(1+0.76\,\chi^2\right)M_B\,.
\end{equation}
In particular, the mean square error of the fit is $\approx0.02$, indicating a very good agreement with the true data points.
\subsection{Compactness} Fig.~\ref{fig:radius} shows the inverse compactness $C^{-1}=R/M$ as plotted in \cite{ryan1997} for a spinning BS. We interpolated the data points with a cubic spline and extracted from the interpolation the values of $C^{-1}$ at $\chi=0$. We find that, at $\chi=0$, the following quadratic fitting formula is a good approximation of the numerical data:
\begin{equation}
    \label{compactness:1}
    \begin{split}
    C^{-1}&=56.3-97.7\left(\frac{M}{M_{\rm max}^{\rm static}}\right)+48.8\left(\frac{M}{M_{\rm max}^{\rm static}}\right)^2\\
    &\approx 7.5+48.8\left(1-\frac{M}{M_{\rm max}^{\rm static}}\right)^2\,,
    \end{split}
\end{equation}
from which we see that $C$ ranges from $\sim0.13$ when $M=M_{\rm max}^{\rm static}$ to $\sim0.034$ when $M=0.02M_B$ (recall that $M_{\rm max}^{\rm static}=0.06 M_B)$.

\subsection{Tidal deformability} 
\begin{figure}[hbt!]
    \centering
    \includegraphics[width=0.4\textwidth]{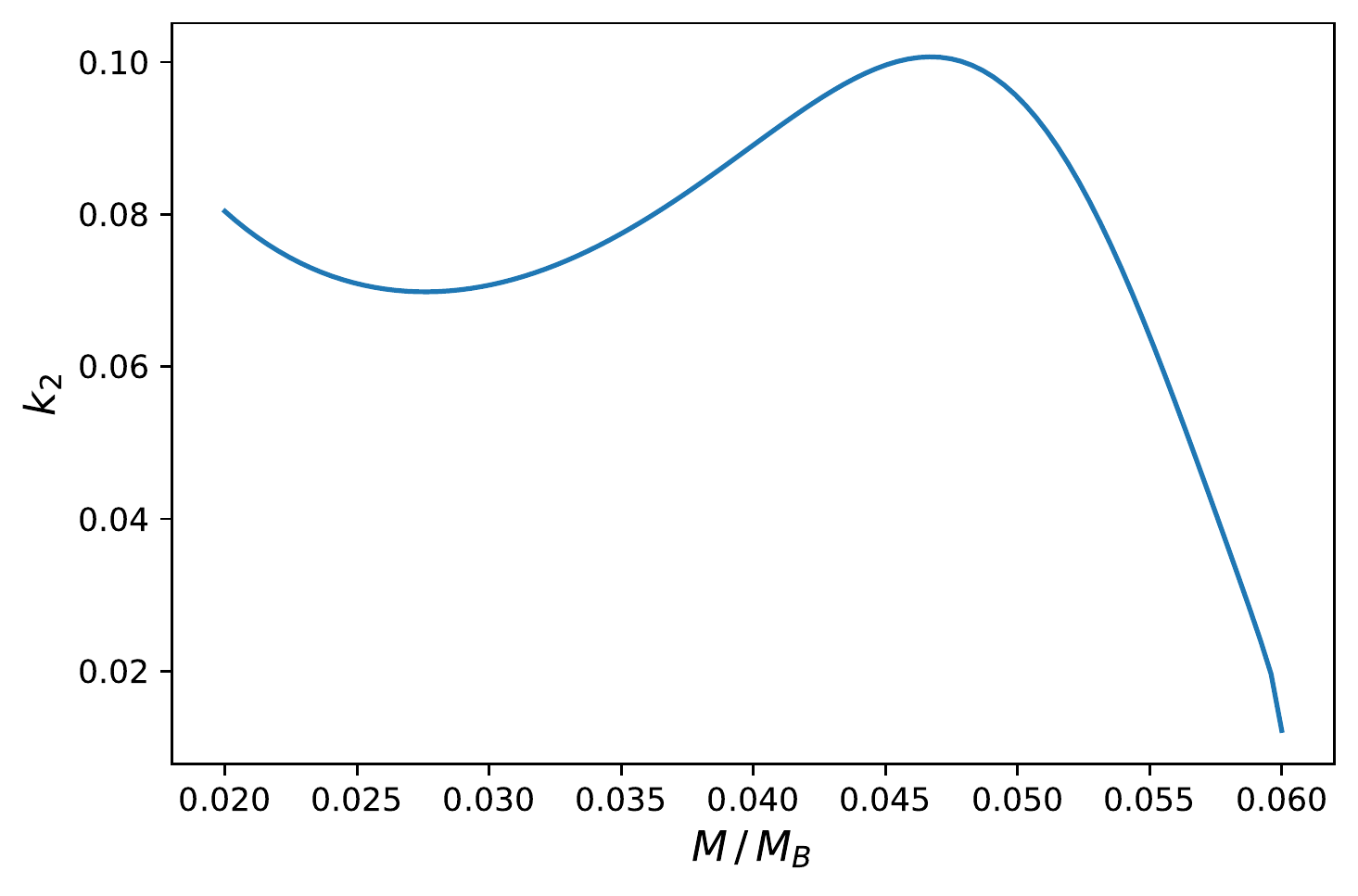}
    \caption{$k_2$ versus $M/M_B$, as resulting from the equation $\Lambda=k_2\times C^5$ and from the Eqs.~\eqref{fit:lambda} and \eqref{compactness:1} for $\Lambda$ and $C$, respectively.}
    \label{fig:k_2}
\end{figure}

\begin{figure}[hbt!]
    \centering
    \includegraphics[width=0.46\textwidth]{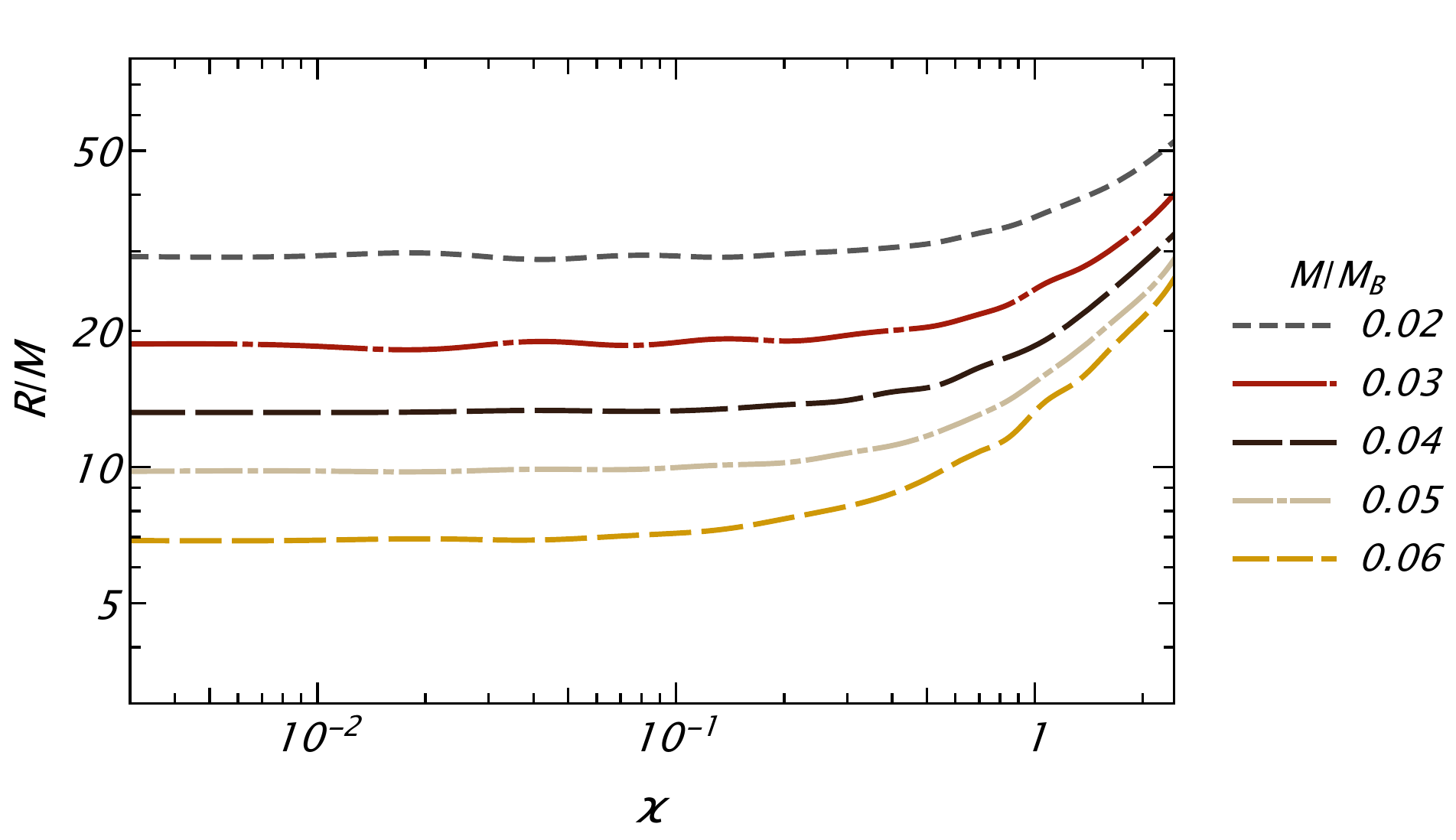}
    \caption{Inverse compactness $C^{-1}=R/M$ of a massive BS. The data were extracted and interpolated from Fig.~2 in Ref.~\cite{ryan1997}.}
    \label{fig:radius}
\end{figure}

We extract the data for the tidal deformability from the fitting formula~\eqref{fit:lambda}, which is obtained from Eq.~(47) in Ref.~\cite{sennett2017} in the limit $\sqrt{\lambda}/\mu^2\gg1$.
Next, we define a new quantity $k_2$ by rescaling $\Lambda$ with $C^5$, $k_2 \equiv \Lambda\, C^5$. Figure~\ref{fig:k_2} shows the profile of $k_2$ as a function of $M/M_B$: we see that approximately $k_2\in(0.01,0.1)$. This, together with the small values that can be attained by the compactness $C$, explains why $\Lambda$ can take large values as $M/M_B$ decreases.
\subsection{Spin-induced quadrupole moment}
\begin{figure}[hbt!]
    \centering
    \includegraphics[width=0.4\textwidth]{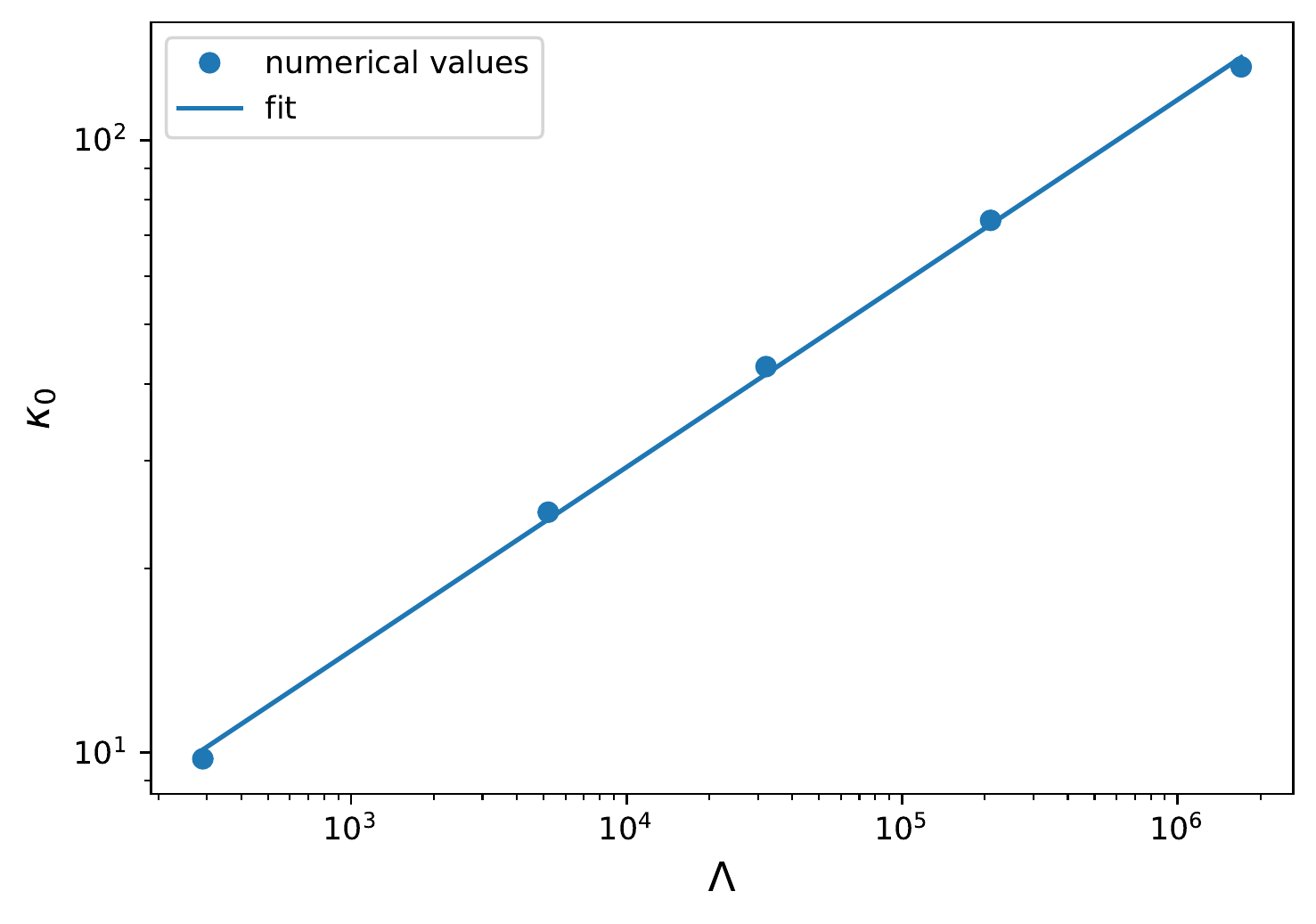}
    \caption{Love-Q relation~\eqref{love:q:1}: $\kappa_0$ versus $\Lambda$ for a slowly rotating BS.}
    \label{fig:love:q}
\end{figure}

We extracted numerical values for $\kappa$ from Fig.~4 in Ref.~\cite{ryan1997} and interpolated them with a cubic spline. Then we extrapolated down to $\chi=0$, thus obtaining the values of $\kappa_0=\kappa(\chi=0)$ corresponding to $M/M_B=0.02,0.03,0.04,0.05,0.06$.
It is intriguing that, as shown in Fig.~\ref{fig:love:q}, $\kappa_0$ and $\Lambda$ obey a simple relation of the form
\begin{equation}
    \log\kappa_0 = 0.61+0.30\left(\log\Lambda\right)\,.
\end{equation}
In particular the mean square error is $\approx0.01$, indicating an excellent agreement.

\bibliography{taylorbs}
\end{document}